\documentclass[aps, pra, reprint, groupedaddress, amsfonts, longbibliography,
               amsmath, amssymb, showpacs, nofootinbib]{revtex4-1}

\usepackage{microtype}
\usepackage{graphicx}
\usepackage{epstopdf}
\usepackage[utf8]{inputenc}
\usepackage[T1]{fontenc}
\usepackage[usenames,dvipsnames]{xcolor}
\usepackage{hyperref}

\newcommand{\bra}[1]{\langle #1|}
\newcommand{\ket}[1]{|#1\rangle}
\newcommand{\braket}[2]{\langle #1|#2\rangle}
\def\idt{{i} \partial _t}
\usepackage{color}% \textcolor{#1}{#2}

\begin{document}
\title{A screened independent atom model for the description of ion collisions from 
atomic and molecular clusters}
\author{Hans J\"urgen L\"udde}
\email[]{luedde@itp.uni-frankfurt.de}
\affiliation{Institut f\"ur Theoretische Physik, Goethe-Universit\"at, D-60438 Frankfurt, Germany} 

\author{Marko Horbatsch}
\email[]{marko@yorku.ca}
\affiliation{Department of Physics and Astronomy, York University, Toronto, Ontario M3J 1P3, Canada}

\author{Tom Kirchner}  
\email[]{tomk@yorku.ca}
\affiliation{Department of Physics and Astronomy, York University, Toronto, Ontario M3J 1P3, Canada}
\date{\today}
\begin{abstract}
We apply a recently introduced model for an independent-atom-like
calculation of ion-impact electron transfer and ionization
cross sections to proton collisions from water, neon, and
carbon clusters.
The model is based on a geometrical interpretation of the cluster cross section as an 
effective area composed of overlapping circular disks that are representative of the atomic contributions. 
The latter are calculated using a time-dependent density-functional-theory-based single-particle
description with accurate exchange-only ground-state potentials.
We find that the net capture and ionization cross sections in p-X$_n$ collisions
are proportional to $n^\alpha$ with $2/3 \le \alpha \le 1$. For capture from water clusters at 100 keV impact
energy $\alpha$ is close to one, which is substantially different
from the value $\alpha=2/3$ predicted by a previous theoretical work based on
the simplest-level electron nuclear dynamics method. 
For ionization at 100 keV and for capture at lower energies we find smaller $\alpha$ values than
for capture at 100 keV. 
This can be understood by considering the magnitude of the atomic cross sections and the
resulting overlaps of the circular disks that make up the cluster cross section in our
model. Results for neon and carbon clusters confirm these trends.
Simple parametrizations are found which fit the cross sections remarkably well and suggest
that they depend on the relevant bond lengths.
\end{abstract}

\maketitle
\section{Introduction}
\label{intro}
Ionization in charged-particle matter interactions is a process of 
relevance to both fundamental and more applied research areas, but is difficult
to describe in quantitative terms if the objects under study are sufficiently complex. 
On the experimental side, challenges associated with preparation and control of
the projectile and target species as well as the detection of multiple reaction products,
possibly in coincidence, have to be addressed.
On the theoretical side, the challenge resides in the 
description of an interacting
few- or many-body system far away from its ground state, a problem that is
straightforward to formulate for (nonrelativistic) Coulomb systems, but hard to solve 
even with present-day supercomputers \cite{gainullin15}.

Time-dependent density functional theory (TDDFT) was 
conceived 
by Erich Runge and Hardy Gross \cite{RG84} with this problem in mind and the objective to
develop a time-dependent description of scattering experiments that would circumvent
the calculation of the many-body wave function \cite{marques04}.
However, applications of the time-dependent Kohn-Sham (TDKS) scheme to collision problems have
remained relatively sparse. This is different from the situation for the somewhat related
problem of ionization in strong 
laser fields (see, e.g., the books \cite{tddft,ullrich} and references therein).  
There are no obvious symmetries in
collisional ionization 
and furthermore for
positively charged projectile ions direct target ionization competes with
electron transfer to bound projectile states.
In the framework of the semiclassical approximation, in which the projectile is assumed
to move on a classical (straight-line) path, projectile-centered states must be augmented by
so-called electron translation factors (ETFs) to account for the relative motion and preserve
Galilean invariance.

These issues have been analyzed in some detail for the 
two-center ion-atom case, often for prototypical one-electron problems such as
the proton-hydrogen collision system \cite{Fritsch91,Bransden92}. Collisions involving helium are perhaps the next
best studied systems, but the vast majority of calculations have been based on simplified descriptions since
explicit solutions of the two-electron time-dependent
Schr\"o\-din\-ger equation are exceedingly difficult and 
computationally costly (see, e.g., reference \cite{Baxter16} and references therein). 
For atoms with more than two electrons, let alone for the atomic
and molecular clusters addressed in the present work, 
they are out of reach.

A popular framework for a simplified treatment of many-electron collision systems is 
the independent electron model (IEM). 
However, 
the most sophisticated IEM variant, the time-dependent Hartree-Fock (TDHF) scheme,
has been applied to only a handful of cases. This is due to difficulties associated with the nonlocal
exchange interaction and, more fundamentally, with the nonlinearity of the TDHF equations,
which manifests itself in the occurrence of fluctuating transition probabilities when
analyzing the TDHF wave function with respect to eigenstates of a static asymptotic
Hamiltonian \cite{kulander82,stich83,gramlich86}. 
The latter problem is known as the TDHF cross channel correlation or projection problem and has also been discussed 
in the context of nuclear reactions \cite{griffin80,alhassid81}.

About 20 years ago we started to look into atomic collisions involving many-electron targets such as
neon and argon atoms using a TDDFT-inspired single-particle description based on 
atomic ground-state DFT potentials \cite{tom97,tom98}. 
The time dependence of these potentials over the course of the collision was
neglected and the projection problem avoided. 
Orbital propagation was achieved using 
the basis generator method (BGM),
a basis set expansion technique built on atomic
orbitals and dynamically adapted pseudostates \cite{BGM99}. The BGM and the more recent two-center version
TC-BGM~\cite{tcbgm} proved capable of accounting for target excitation, electron transfer, and ionization
channels in ion-atom collision problems over a wide range of collision energies (see, e.g., 
references \cite{Kirc04,leung17} and references therein).
 
Subsequently, we amended the {\it no-response} 
approximation of frozen ground-state potentials by simple
response models which did not increase the computational burden significantly
and allowed us to analyze the projection problem and
study dynamical potential effects in a semi-quantitative 
way \cite{tom00c}. We found that response has the tendency to lower probabilities for total
electron removal (the sum of electron transfer to the projectile and
direct target ionization) at low to intermediate collision energies.
As soon as the projectile speed is significantly
larger than the average orbital velocity of a given target electron,
that electron becomes insensitive to 
time-dependent changes in the interelectronic potential simply because ionization happens
too rapidly. As a consequence, response model cross sections  
approach no-response results towards high collision energies.
Also, response effects turned out ot be generally small for proton impact on first-row elements. 
This is so because
multiple-electron removal is a weak process in these collisions and our model is
designed in such a way that dynamical screening becomes appreciable only after one electron is
removed on average. This choice was motivated by the success of so-called frozen TDHF 
calculations in studies concerned with the (photo-) ionization of a single electron \cite{kulander88}.

Collisions of projectile ions (with or without projectile electrons) from small molecules, 
such as $\rm H_2O$ \cite{hjl09,mitsuko12}, or $\rm CH_4$ \cite{arash17}
were treated within a framework where simple self-consistent field wave functions were projected onto atomic orbitals
calculated in DFT. These orbitals were then evolved using the TC-BGM and transition amplitudes were calculated on the basis of
interpreting Kohn-Sham determinants. For biomolecules and clusters this approach is not suitable. Direct implementations of
TDDFT equations for ion collisions with small molecules were reported by other groups \cite{Hong16,quashi17}.
For larger systems a few calculations based on first-principles approaches have been 
carried out \cite{Privett14,Montabonel16,Covington17}, 
but most of the available cross section results for electron transfer and target ionization
have been obtained using simplified and classical 
models~(see, e.g., references~\cite{Cappello08,Lekadir09,Champion12,Devera13,Sarkadi15} and
references therein).
  
Given this situation
we recently introduced an indepen\-dent-atom-model (IAM) description
to deal with complex multicenter collision systems 
on the basis of atomic no-response TC-BGM calculations \cite{hjl16}. 
The simplest realization of the IAM is Bragg's additivity rule (IAM-AR)
according to which a net cross section for a complex target such as a molecule or cluster
is obtained from adding up atomic net cross sections for all atoms that make up the system.
Our model goes beyond the IAM-AR by associating
the atomic cross sections in the AR sum 
with weight factors.
The latter are determined from a geometrical interpretation of a cross 
section as an effective area using the following procedure.
First, each atom in a given
target is surrounded by a sphere of a radius representative of the atomic cross section 
for either net electron transfer to the projectile or to the continuum (for brevity referred to as
net capture and net ionization in the following). Secondly,
the resulting three-dimensional structure of overlapping spheres is projected
on a plane which is perpendicular to the projectile
beam axis.
In the last step,  
the effective area in that plane is taken as 
the cross section for net capture or net ionization of the system in that particular
geometry. An orientation average is calculated to make contact with experimental
data for randomly oriented molecules or clusters.
We refer to the model as IAM-PCM, since the effective cross-sectional area,
and by extension the weight factors attached to the atomic contributions in a given
orientation,
are calculated using a pixel counting method (PCM).

The IAM-PCM was successfully applied to a number of collision systems involving proton
projectiles and molecular targets such as CO, H$_2$O, and C$_4$H$_4$N$_2$O$_2$ (uracil).
It was demonstrated that IAM-AR cross sections for net capture and ionization
are reduced substantially and agreement with experimental data is improved in regions in
which the atomic cross section contributions are large and the overlap effects significant \cite{hjl16}. 

In this work, we use the IAM-PCM to calculate net capture and ionization
cross sections in proton collisions with water, neon, and carbon clusters comprising
systems with hydrogen bonds, van der Waals bonds, and covalent bonds.
We begin in section~\ref{sec:bgm} with a discussion of the atomic ingredients,
i.e., the solution of the (approximate) ion-atom TDKS equations using
the TC-BGM (section~\ref{sec:bgm1}),
the calculation of cross sections for net capture and ionization (section~\ref{sec:observables}),
and 
results for the p-H, p-C, p-O, and p-Ne systems (section~\ref{sec:bgm2}).
This is followed by a description of the IAM-PCM in section~\ref{sec:pcm}.
Results for the proton-cluster collision systems are presented in section~\ref{sec:results} and
the paper ends with a few concluding remarks in section~\ref{sec:conclusions}.
Atomic units, characterized by $\hbar=m_e=e=4\pi\epsilon_0=1$, are used unless otherwise stated.

\vspace{1\baselineskip}

\section{The basis generator method for ion-atom collisions}
\label{sec:bgm}

\subsection{Solution of the single-particle equations}
\label{sec:bgm1}
The TDKS scheme was anticipated by Runge and Gross in their original 1984 work \cite{RG84} and
put on firm grounds by van Leeuwen in 1999 \cite{leeuw99}. For a thorough discussion of the
foundational theorems of TDDFT
we refer the reader to the books \cite{tddft,ullrich} and references therein.

For an $N$-electron ion-atom collision problem in the semiclassical approximation the TDKS
equations can be written in the form
\begin{eqnarray}
\idt \psi_i({\bf r},t) &=&  \left(-\frac{1}{2}\nabla^2 - \frac{Z_T}{r_T}
         - \frac{Z_P}{r_P} + v_{ee}[n]({\bf r},t)\right)  \psi_i({\bf r},t) , \nonumber \\
    && \quad i=1,\ldots,N  , 
\label{eq:tdks}
\end{eqnarray}
where $\mathbf r$, $\mathbf r_T$, and $\mathbf r_P$ denote the electronic position vector
with respect to the center of mass (CM), the target, and
the projectile, respectively, and
$Z_T$ and $Z_P$ are the charge numbers of the nuclei.
We assume the projectile to follow a straight-line path ${\bf R}(t)={\bf r}_T-{\bf r}_P=(b,0,vt)$ 
characterized by the impact parameter $b$
and the constant speed $v$. 

The effective electron-electron
potential $v_{ee}$ in equation~(\ref{eq:tdks}) 
is a functional of the density $n$ and
can be split into the usual Hartree, exchange, and correlation contributions.
In the no-response approximation for the problem of a bare projectile ion impinging on an atomic target,
$v_{ee}$ is given by a (spherically-symmetric) ground-state DFT potential. More specifically, we 
use Hartree-exchange potentials obtained from the exchange-only version of the optimized potential method
(OPM) \cite{talman76,ee93} for the neutral ($N=Z_T$) carbon, oxygen, and neon atoms of interest in the present study and neglect correlation
effects. An important feature of the OPM potentials is their complete cancellation of self-interaction
contributions contained in the Hartree potential such that the correct asymptotic behaviour
\begin{equation}
    v_{ee}^{\rm OPM}(r_T) \stackrel{r_T \rightarrow \infty}{\rightarrow } 
   \frac{N-1}{r_T}
\label{tail0}
\end{equation}
is ensured. This is crucial for a proper description of target electron removal~\cite{tom97,tom98}.

The approximate TDKS equations (\ref{eq:tdks}) with the ground-state potential 
$v_{ee}^{\rm OPM}$ are propagated using the TC-BGM, which, like any
basis-expansion technique, assumes that the solutions can be represented in terms of a finite set of states.
The TC-BGM set includes $N_T$ atomic states on the target center, i.e., bound eigenstates of
\begin{eqnarray}
\hat h_T &=& -\frac{1}{2}\nabla^2 - \frac{Z_T}{r_T} + v_{ee}^{\rm OPM}(r_T) \\
         &\equiv& -\frac{1}{2}\nabla^2 + V_T ,
\label{eq:ht}
\end{eqnarray}
a set of $N_P$ eigenstates of the projectile Hamiltonian
\begin{equation}
\hat h_P = -\frac{1}{2}\nabla^2 - \frac{Z_P}{r_P} \equiv -\frac{1}{2}\nabla^2 + V_P 
\label{eq:hp}
\end{equation}
to describe capture, 
and a set of pseudostates which overlap with the continuum.
It is the specific choice of the latter that distinguishes the TC-BGM from other coupled-channel methods
for atomic collisions.
The guiding idea is to span a subspace of Hilbert space which dynamically adapts to
the time evolution of the system in such a way that 
couplings to the complementary space are small and can be neglected without introducing 
significant errors. The benefit of using time-dependent basis states is that one can hope for
reasonable convergence without having to include a very large number of states.

It was shown on theoretical grounds \cite{BGM99} and demonstrated in a number of practical 
applications~(see, e.g., references \cite{Kirc04,leung17} and references therein)
that good convergence can be achieved by using basis functions of the form
\begin{eqnarray}
\chi_{j}^J (\mathbf{r},t)  &=& [W_P(r_P)]^{J} \phi_j^0(\mathbf{r},t) \label{eq:hierarchy}\\
W_{P}(r_P)  &=&  \frac{1}{r_P}\left(1-{\rm e}^{-r_P}\right) \label{eq:wp} \\
  \phi_j^0({\bf r},t) & = &
\left\lbrace
\begin{array}{ll}
\phi_j({\bf r}_T) \exp(i {\bf v}_T \cdot {\bf r} ) &  \mbox{if}\;\; j\le N_T \\
  \phi_j({\bf r}_P) \exp(i {\bf v}_P \cdot {\bf r}) &  \mbox{if}\;\; N_T < j \le N_T+N_P  ,
\end{array}
\right.
\label{eq:genbgm}
\end{eqnarray}
where ${\bf v}_T$ and ${\bf v}_P$ denote the (constant) velocities of the target and projectile
with respect to the CM, and the functions $\phi_j$  on the right hand side of equation (\ref{eq:genbgm})
satisfy stationary eigenvalue equations for $\hat h_T$ ($j\le N_T$) and 
$\hat h_P$ ($j> N_T$) in the (moving)
target and projectile reference frames, respectively.
The phase factors are ETFs which ensure Galilean invariance.
Acting with spatial and time derivative operators on them leads
to the modified eigenvalue equations
\begin{equation}
(\hat h_{T,P} -\idt) \ket{\phi_j^0}  =  g_j \ket{\phi_j^0}
\end{equation}
with
\begin{equation}
g_j = \varepsilon_j + \frac{v_{T,P}^2}{2}
\end{equation}
and atomic energy eigenvalues $\varepsilon_j$ for the
target and projectile orbitals $\phi_j^0({\bf r},t) = \braket{{\bf r}}{\phi_j^0}$ 
in the CM reference frame.

Expanding the TDKS orbitals in this non-orthogonal, time-dependent TC-BGM basis
\begin{equation}
 \psi_i({\bf r},t) = \sum_{j,J} c_{j,J}^i(t) \chi_j^J (\mathbf{r},t)
\end{equation}
turns the single-particle equations into a set of coupled  equations for the expansion coefficients
\begin{equation}
       i \sum_{j,J}   S_{kj}^{KJ}(t) \dot{c}_{j,J}^i(t)  =
          \sum_{j,J}
            M_{kj}^{KJ}(t) c_{j,J}^i(t)
\label{eq:coupled}
\end{equation}
with overlap
\begin{equation}
      S_{kj}^{KJ} = 
          \braket{kK}{jJ}   
\label{eq:skj}
\end{equation}
and interaction 
\begin{equation}
      M_{kj}^{KJ} = 
       \bra{kK} \hat{h}_{T}  + V_P - \idt
          \ket{jJ}   
\label{eq:mkj}
\end{equation}
matrix elements. In equations (\ref{eq:skj}) and (\ref{eq:mkj}) we have used the short-hand notation
\begin{equation}
\ket{jJ} = W_P^J\ket{j0}  
\end{equation}
for the BGM basis states, i.e., the functions $\chi_j^J (\mathbf{r},t) = \braket{{\bf r}}{jJ}$.

The calculation of the matrix elements proceeds in several steps. First, the interaction matrix
elements (\ref{eq:mkj}) are rewritten by using similar arguments as in references~\cite{hjl96}
and \cite{mitsuko12} to arrive at
\begin{eqnarray}
M_{kj}^{KJ} & = & \bra{kK}\frac{KJ}{2}\left(\frac{\nabla W_P}{W_P}\right)^2 + \frac{K}{K+J}V_{\bar j} + 
      \frac{J}{K+J}V_{\bar k} \ket{jJ} \nonumber \\
 & & \mbox{} - \frac{J}{K+J}\idt\braket{kK}{jJ} + \frac{Kg_j + J g_k}{K+J}  \braket{kK}{jJ} ,
\label{eq:mkjex}
\end{eqnarray}
where for $j\le N_T$ we set $V_{\bar j} = V_P$ and $V_j = V_T$,
while for $j > N_T$ we set $V_{\bar j} = V_T$ and $V_j = V_P$. 
In contrast to equation (\ref{eq:mkj}) the equivalent form (\ref{eq:mkjex}) does not involve
derivatives of basis functions.

In a second step, the set of TC-BGM pseudostates $\{\ket{jJ}, J>0 \}$ is orthogonalized to
the generating two-center basis $\{\ket{j0}  \}$
to separate the ionized and bound parts of the TDKS orbitals.
Finally, an LU decomposition 
is carried out to turn the basis into a completely orthonomalized set of states and 
the coupled-channel equations (\ref{eq:coupled}) into the form
\begin{equation}
   i \dot{{\bf d}^i}  = \tilde{\mbox{\sf M}} {\bf d}^i ,
\label{eq:matrix}
\end{equation}
in which $\tilde{\mbox{\sf M}}$ is the transformed interaction matrix
and ${\bf d}^i$ the transformed expansion coefficient vector of the $i$-th TDKS orbital.
The set of matrix equations (\ref{eq:matrix}) is solved using standard methods~\cite{odepack}.

\begin{figure*}
\begin{center}
\resizebox{0.49\textwidth}{!}{%
\includegraphics{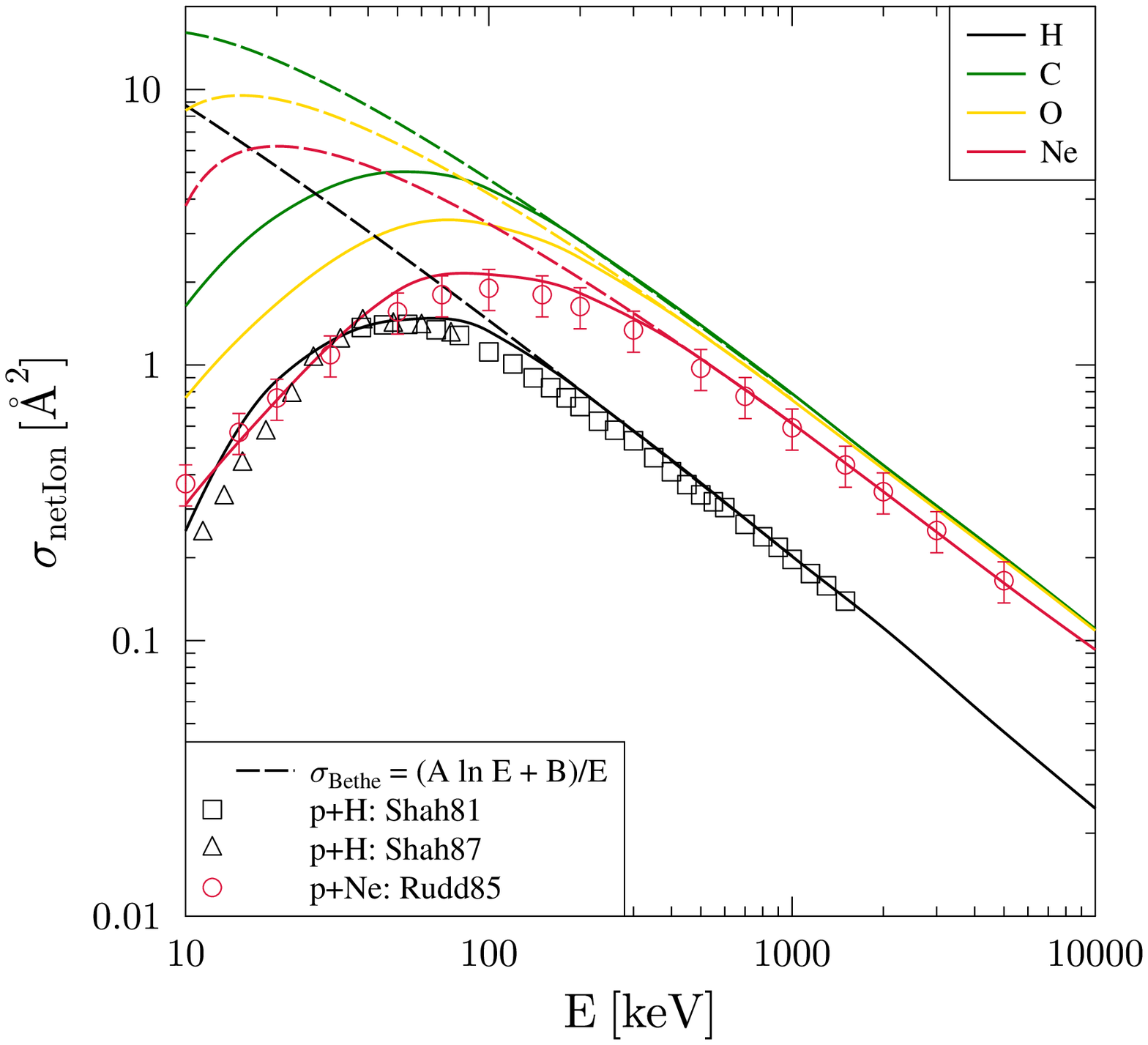}
}
\resizebox{0.49\textwidth}{!}{%
\includegraphics{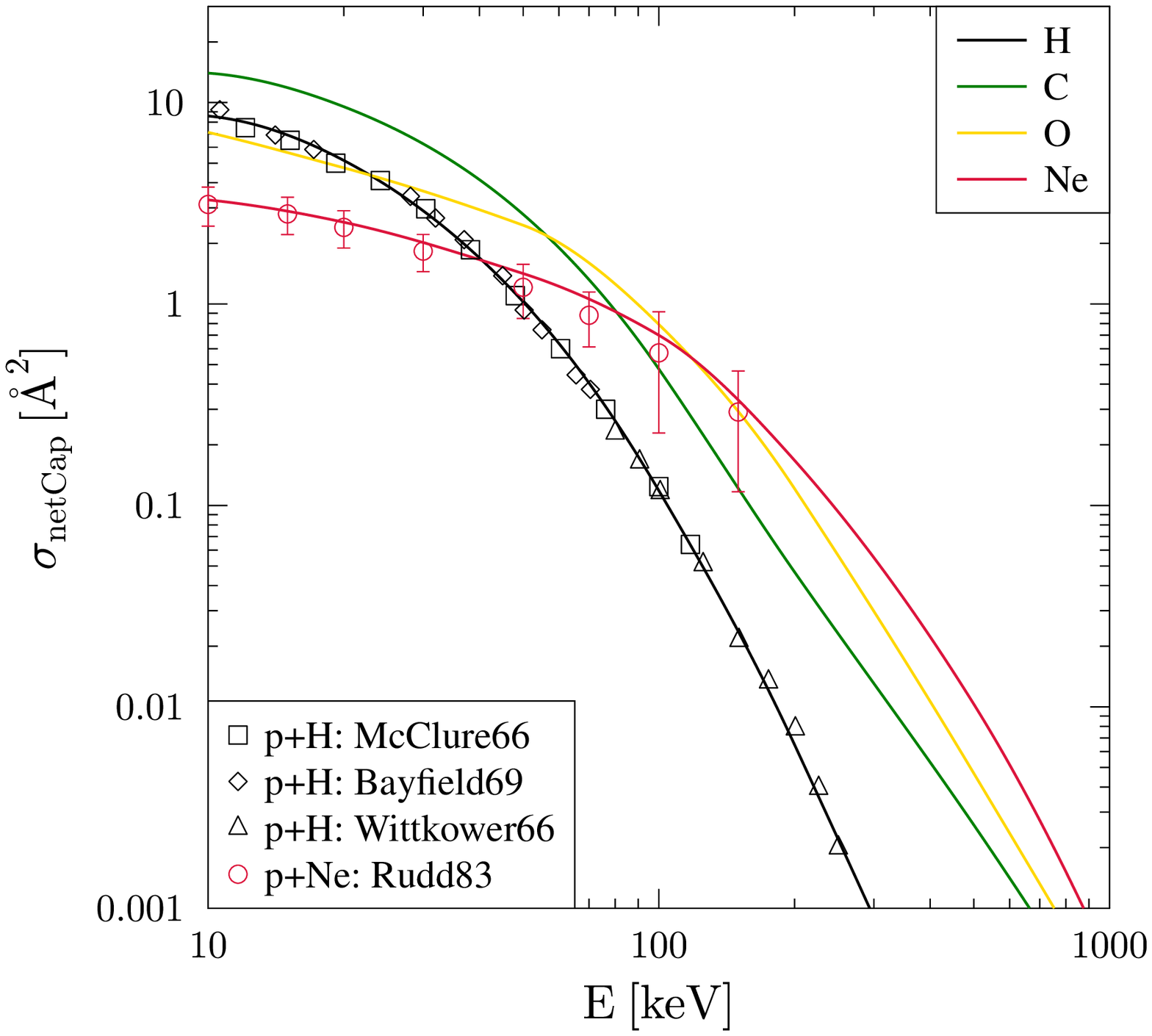}
}

\vspace{-2.5\baselineskip}
\caption{Total cross sections for net ionization (left panel) and net capture (right panel)
in p-H, p-C, p-O, and p-Ne collisions as functions of impact energy.
Experiments: Shah81~\cite{Shah81}, Shah87~\cite{Shah87}, Rudd85~\cite{Rudd85a}, 
McClure66~\cite{McClure66}, Bayfield69~\cite{Bayfield69}, Wittkower66~\cite{Wittkower66},
Rudd83~\cite{Rudd83}. 
For the p-H system the reported experimental uncertainties are
below 10\% and the error bars are smaller than
the size of the symbols.}
\label{fig:net-ao}
\end{center}
\end{figure*}

\subsection{Calculation of net cross sections}
\label{sec:observables}
The atomic contributions used in the IAM-AR and IAM-PCM are cross sections 
for net capture and net ionization. They are
calculated, exploiting cylindrical symmetry, via
\begin{equation}
      \sigma^{{\rm net} \, x} = 2\pi \int_0^\infty bP^{{\rm net} \, x}(b) db ,
\label{eq:tcs}
\end{equation}
where $x$ denotes capture ($x=$ cap) or ionization ($x=$ ion) and $P^{{\rm net} \, x}$ is
the corresponding (impact-parameter-dependent) net electron number.
Provided that at an asymptotic time $t_f$ after the collision the one-particle
density $n$ can be split into non-overlapping contributions associated with
electrons captured by the projectile ($P$), promoted to the continuum ($C$) and retained
by the target ($T$), one can write for the total electron number \cite{Luedde03}
\begin{equation}
     N = \int_P n({\bf r},t_f) d^3 r + \int_C n({\bf r},t_f) d^3 r + \int_T n({\bf r},t_f) d^3 r ,
\label{eq:volume}
\end{equation}
where the integrals are over (non-overlapping) $P$, $C$, and $T$ subspaces, and identify
\begin{eqnarray}
   P^{{\rm net \, cap} } &=&  \int_P n({\bf r},t_f) d^3 r , \label{eq:netcap}\\
   P^{{\rm net \, ion}}  &=&  \int_C n({\bf r},t_f) d^3 r .
\label{eq:netion}
\end{eqnarray}
Equations (\ref{eq:netcap}) and (\ref{eq:netion}) show that
net electron numbers, and as a consequence of equation (\ref{eq:tcs}) net cross sections as well, 
are explicit density functionals.
This makes them convenient observables in TDDFT-based studies: The only fundamental
approximation involved in a TDDFT net cross section calculation is the choice made for the
TDKS potential. If one wishes to calculate a cross section that corresponds to a coincident
measurement of single or multiple capture and ionization, one faces the
additional challenge that the exact density dependence of the observables is not known
and additional approximations are required \cite{Baxter16,Luedde03}.

We conclude this section by noting that instead of integrating the electron density over
subspaces of ${\cal R}^3$ we use the TC-BGM basis representation to calculate net capture and ionization
directly from the asymptotic expansion coeffients of equation (\ref{eq:matrix})
\begin{eqnarray}
   P^{{\rm net \, cap} } &=& \sum_{i=1}^N \sum_{k}^P |d_k^{\, i}(t_f)|^2  , \label{eq:netcap2}\\
   P^{{\rm net \, ion}}  &=& N -  P^{{\rm net \, cap} } - \sum_{i=1}^N \sum_{k}^T |d_k^{\, i}(t_f)|^2   .
\label{eq:netion2}
\end{eqnarray}
If the sums over $k$ include all appreciably populated bound projectile ($P$) and target ($T$) states
and provided the above-mentioned condition of non-overlapping $P$, $T$, and $C$ components is fulfilled,
the channel and real-space representations
of $P^{{\rm net \, cap} }$ and $P^{{\rm net \, ion} } $ are equivalent. 
 
\subsection{Sample results}
\label{sec:bgm2}
In Figure~\ref{fig:net-ao} we show no-response TC-BGM
net ionization and net
capture cross section results for the proton-atom collision systems of interest in this
work: p-H, p-C, p-O, and p-Ne. The p-H system in particular has been studied extensively
over the years and many sets of theoretical results have been reported.
Figure~\ref{fig:net-ao} does not provide comparisons with those
previous calculations, since a review of the current status of atomic cross section calculations
is outside the scope of this article.
The purpose of Figure~\ref{fig:net-ao} is limited 
to an illustration of the level of accuracy and the asymptotic behaviour obtained
in the (no-response) TC-BGM framework. To this end,
experimental data for p-H and p-Ne, the
only systems for which direct measurements of net ionization and capture cross
sections are available, 
and fits of the asymptotic Bethe-Born ionization
cross section formula \cite{Bethe30,Inokuti71}
\begin{equation}
\sigma_{\rm Bethe} = \frac{A \ln E +B}{E} ,
\label{eq:bb}
\end{equation}
in which $E$ is the projectile energy and $A$ and $B$ are treated as fit parameters,
are included.
For a broader discussion of p-H cross section results we refer the reader
to the recent work~\cite{Abdura16}. The p-O and p-Ne systems were studied in our
previous papers \cite{tom00b} and \cite{tom00a}, respectively. 

For the various atomic targets
we included in the basis all atomic orbitals of the $KLMN$ shells of both
projectile and target  
plus sets of 73--111 pseudostates constructed according to 
equations~(\ref{eq:hierarchy}) and~(\ref{eq:wp}).
The Bethe-Born cross sections were obtained by fitting the parameters $A$ and $B$ of
equation~(\ref{eq:bb})
to the current TC-BGM results at high energies using the Fano representation,
in which $E\sigma^{\rm net \, ion}$ is plotted against $\ln(E)$ (using appropriate units) \cite{Inokuti71}.
For the p-H system the fitted parameters are consistent with the
values that can be deduced from Bethe's original work \cite{Bethe30}.

Obviously, the agreement of the TC-BGM results with the experimental data and the Bethe-Born prediction
at high $E$ is very good. It is interesting to see that the net ionization cross sections for 
p-C and p-O do not
only agree in shape, but also in magnitude in this region. For the oxygen case
we found excellent agreement with experimental data for equivelocity electron impact corresponding to
$E\ge 200$ keV \cite{tom00b}, which confirms
the validity of first-order perturbation theory.
Furthermore, within 10\% accuracy 
the high-energy p-C and p-O results 
are four times larger than the p-H ionization cross section and, as found in additional
calculations (not included in Figure~\ref{fig:net-ao}), 
they also coincide (within 10\%) with results for p-N collisions. This
implies that for a large class of biomolecules consisting of H, C, N, and O atoms the IAM-AR
will predict very simple scaling relations. We found, somewhat surprisingly, that the same relations
hold within the IAM-PCM described in the next section. 
An analysis of these scaling relations will be presented in a future publication 
focusing on ion-biomolecule collisions.

\section{A pixel counting method for screened independent atom model calculations}
\label{sec:pcm}
The IAM-PCM is best explained by way of an example. Consider net capture in p-H$_2$O collisions
at relatively low impact energy $E$. The ingredients of the IAM are the net capture cross
sections for p-H and p-O collisions. These cross sections are assigned radii according to
\begin{equation}
   r_j = [\sigma_j^{{\rm net} \, cap}/\pi]^{1/2}  ,
\label{eq:radius}
\end{equation}
where $j=1,2,3$ enumerates the atoms.  
We place the $L=3$ atomic nuclei at their
equilibrium positions in ground-state H$_2$O and  
surround each of them by a sphere of radius $r_j$. 
The impinging projectile then encounters an object made up of overlapping spheres 
and an effective cross-sectional area is determined by projecting that object on
the plane that is perpendicular to the projectile beam.

\begin{figure}
\begin{center}
\resizebox{0.22\textwidth}{!}{%
\includegraphics{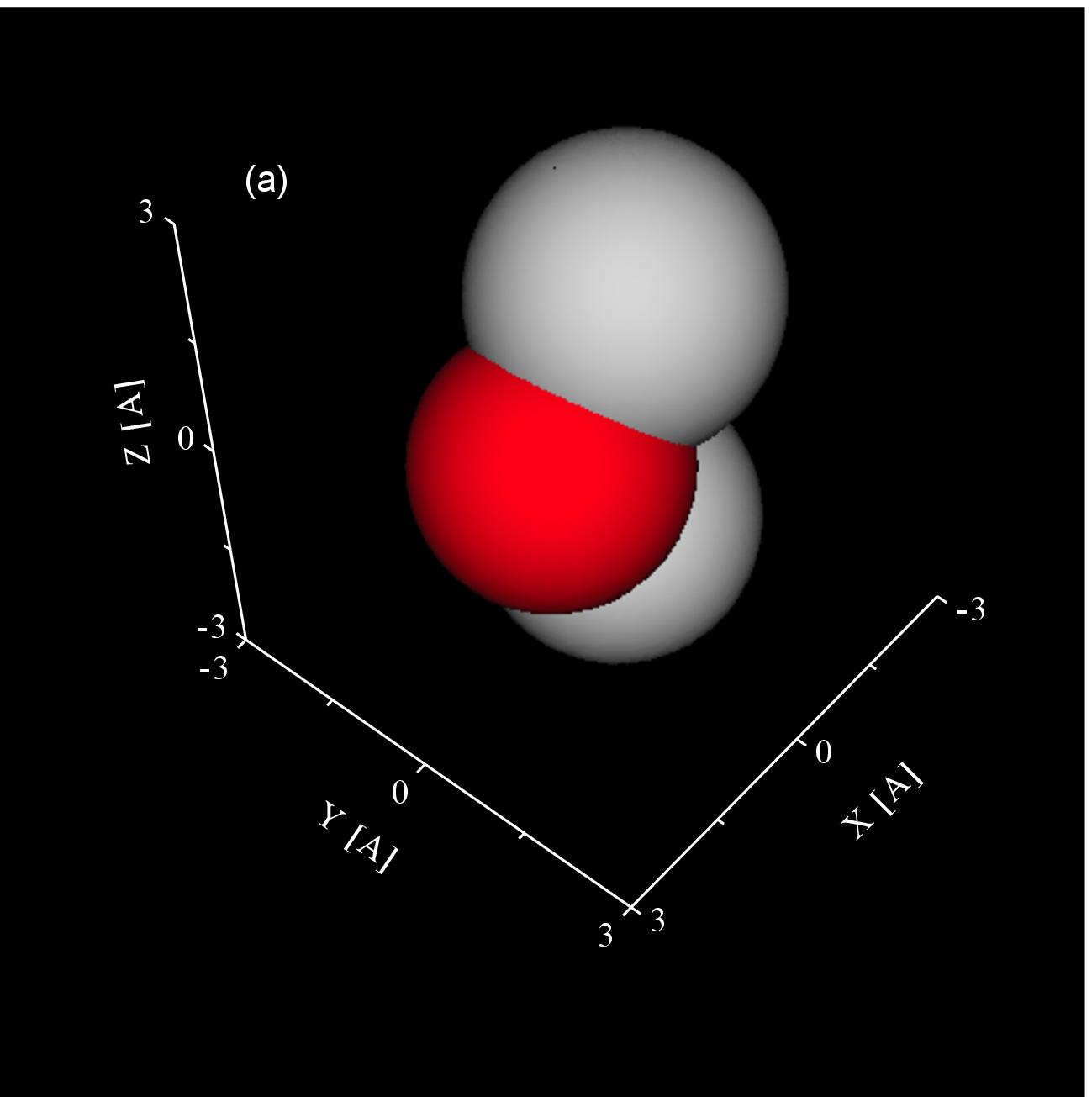}
}
\resizebox{0.235\textwidth}{!}{%
\includegraphics{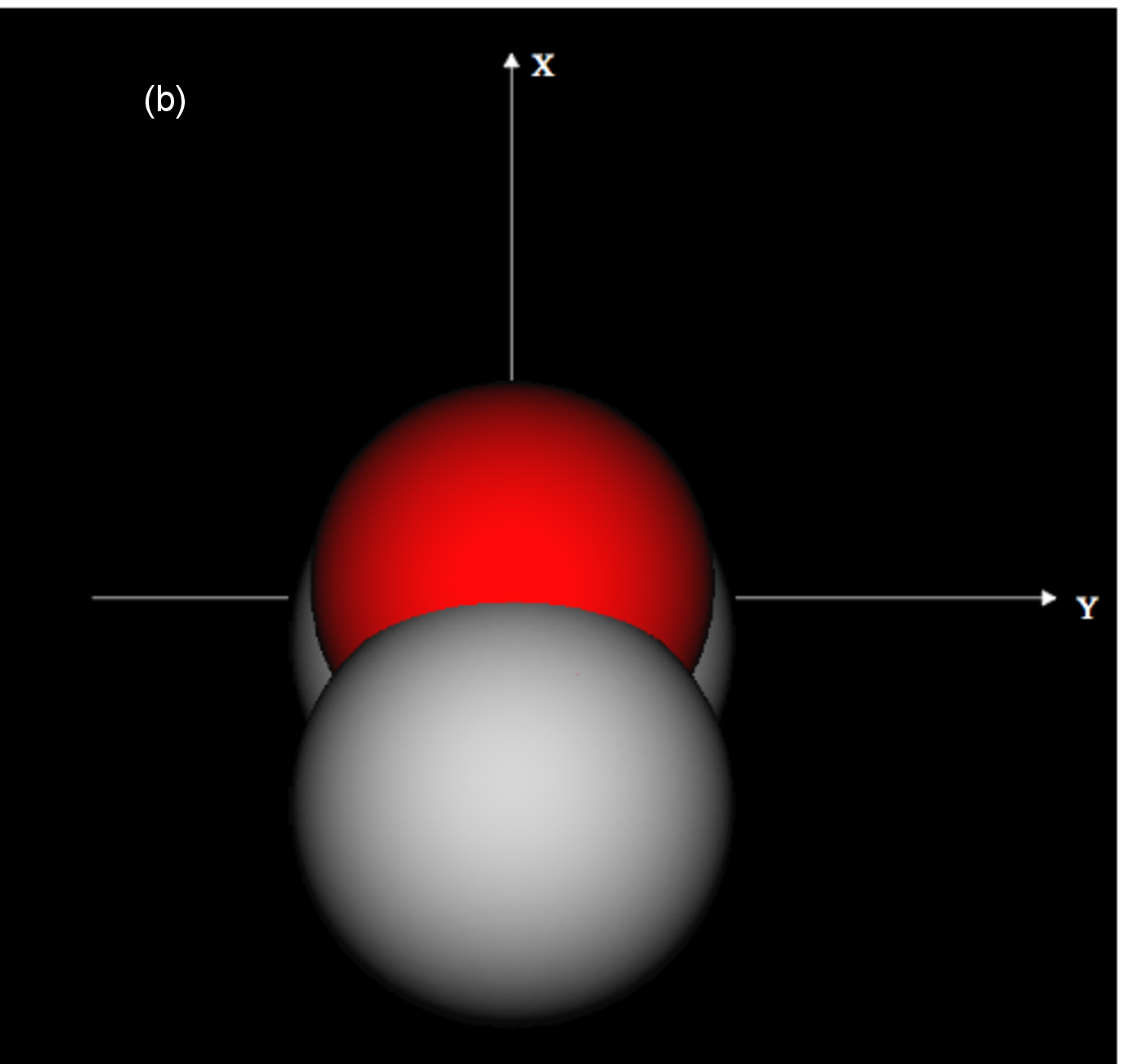}
}
\caption{Net capture 
in p-H$_2$O collisions at $E=10$ keV:
(a) three-dimensional image and (b) projection on
the $x$-$y$ plane. The radii of the spheres and circular disks are determined
acoording to equation~(\ref{eq:radius}).}
\label{fig:10kev}
\end{center}
\end{figure}

Figure~\ref{fig:10kev}a displays the overlapping spheres for capture 
at $E=10$ keV. It is important to keep in mind that the object shown is not a model
of the water molecule, but a three-dimensional representation of net capture. 
A projectile approaching the molecule from a given direction will 'see' the projected
cross-sectional area as in classical scattering from
superimposed hard spheres.
Figure~\ref{fig:10kev}b shows this projection for
projectile impact along the $z$-direction of the coordinate system used. The effective
area, i.e., the molecular net cross section, can be represented as a weighted sum
of atomic cross sections
\begin{equation}
 \sigma_{\rm mol}^{{\rm net} \, x}(E,\alpha,\beta,\gamma) = 
  \sum_{j=1}^L s_j^x(E,\alpha,\beta,\gamma)  \sigma_j^{{\rm net} \, x} (E) 
\label{eq:scar-pcm}
\end{equation}
with weight factors $0\le s_j^x\le 1$ and
the Euler angles
$\alpha,\beta,\gamma$ 
which characterize the orientation of the molecule.
The notation used in equation~(\ref{eq:scar-pcm}) shows the dependencies of the
various quantities and indicates that we use the
prescription for both capture and ionization.
We note in passing that the screening corrected additivity rule (SCAR) for
electron-molecule scattering is based on similar ideas and uses a similar equation, but with
orientation-independent
weight factors that are obtained from a heuristic recurrence relation \cite{Blanco2003}.

\begin{figure}
\begin{center}
\resizebox{0.22\textwidth}{!}{%
\includegraphics{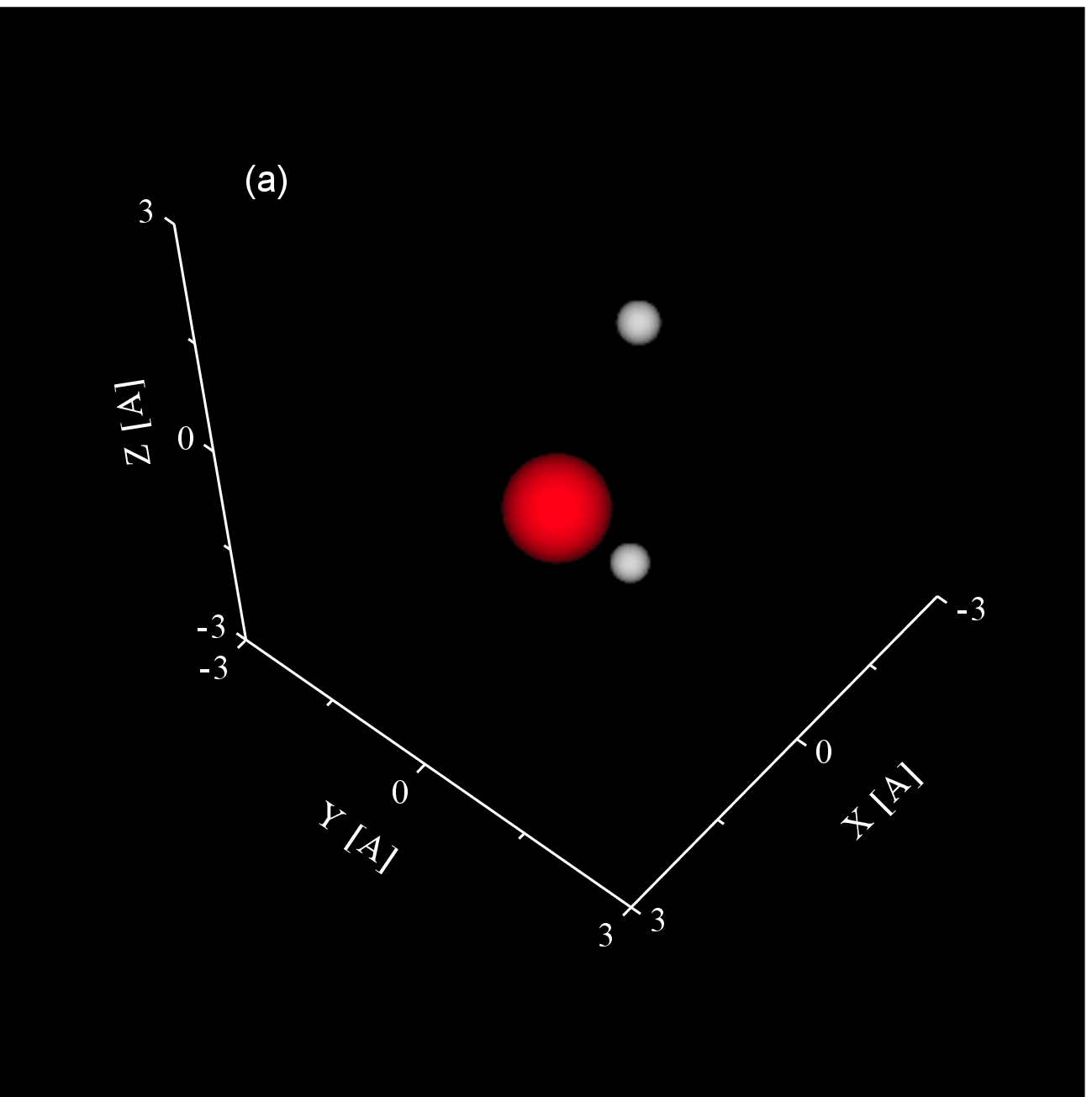}
}
\resizebox{0.23\textwidth}{!}{%
\includegraphics{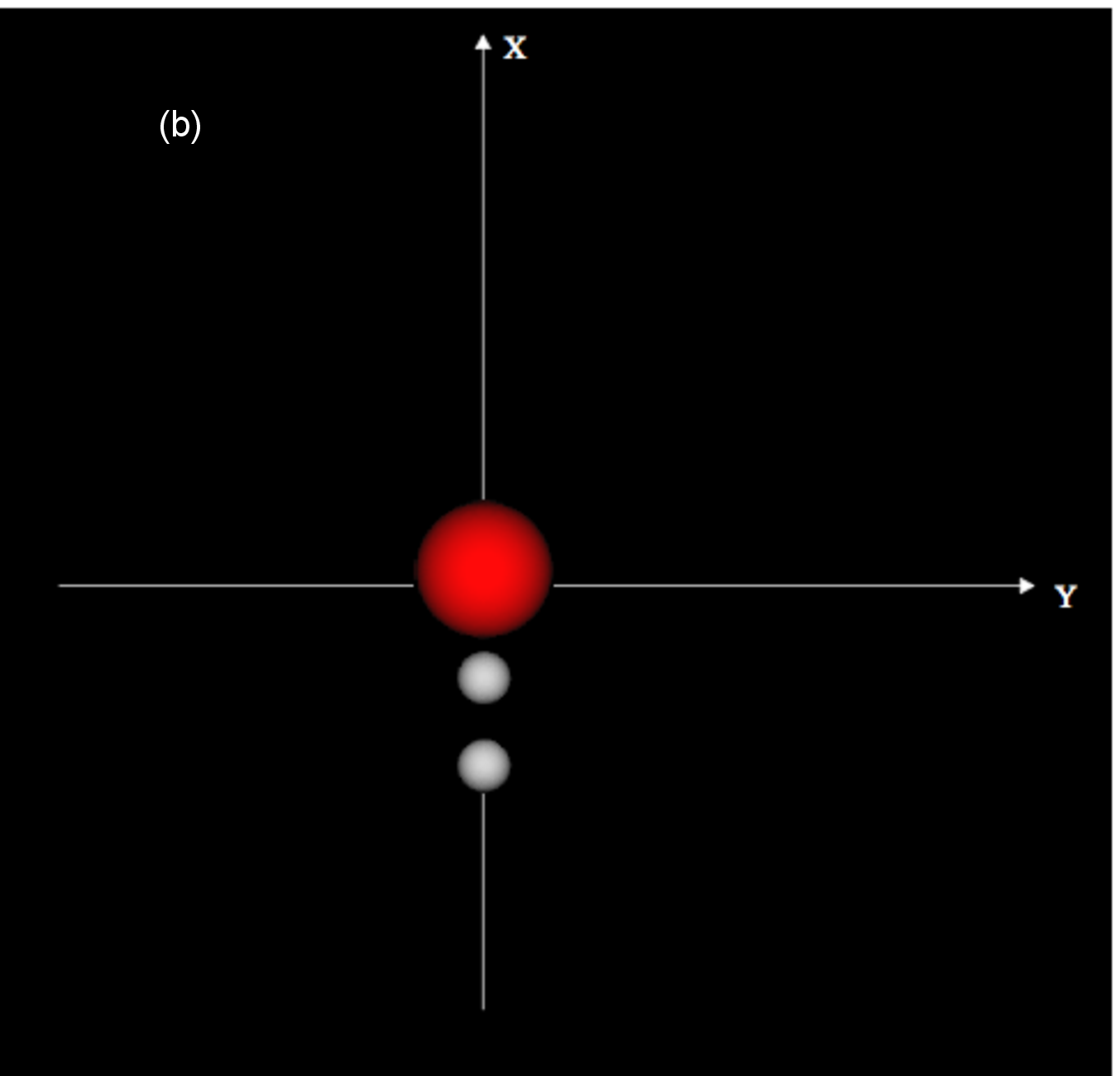}
}
\caption{Net capture 
in p-H$_2$O collisions at $E=100$ keV:
(a) three-dimensional image and (b) projection on
the $x$-$y$ plane. The radii of the spheres and circular disks are determined
acoording to equation~(\ref{eq:radius}).}
\label{fig:100kev}
\end{center}
\end{figure}

\begin{figure*}
\begin{center}
\resizebox{0.49\textwidth}{!}{%
\includegraphics{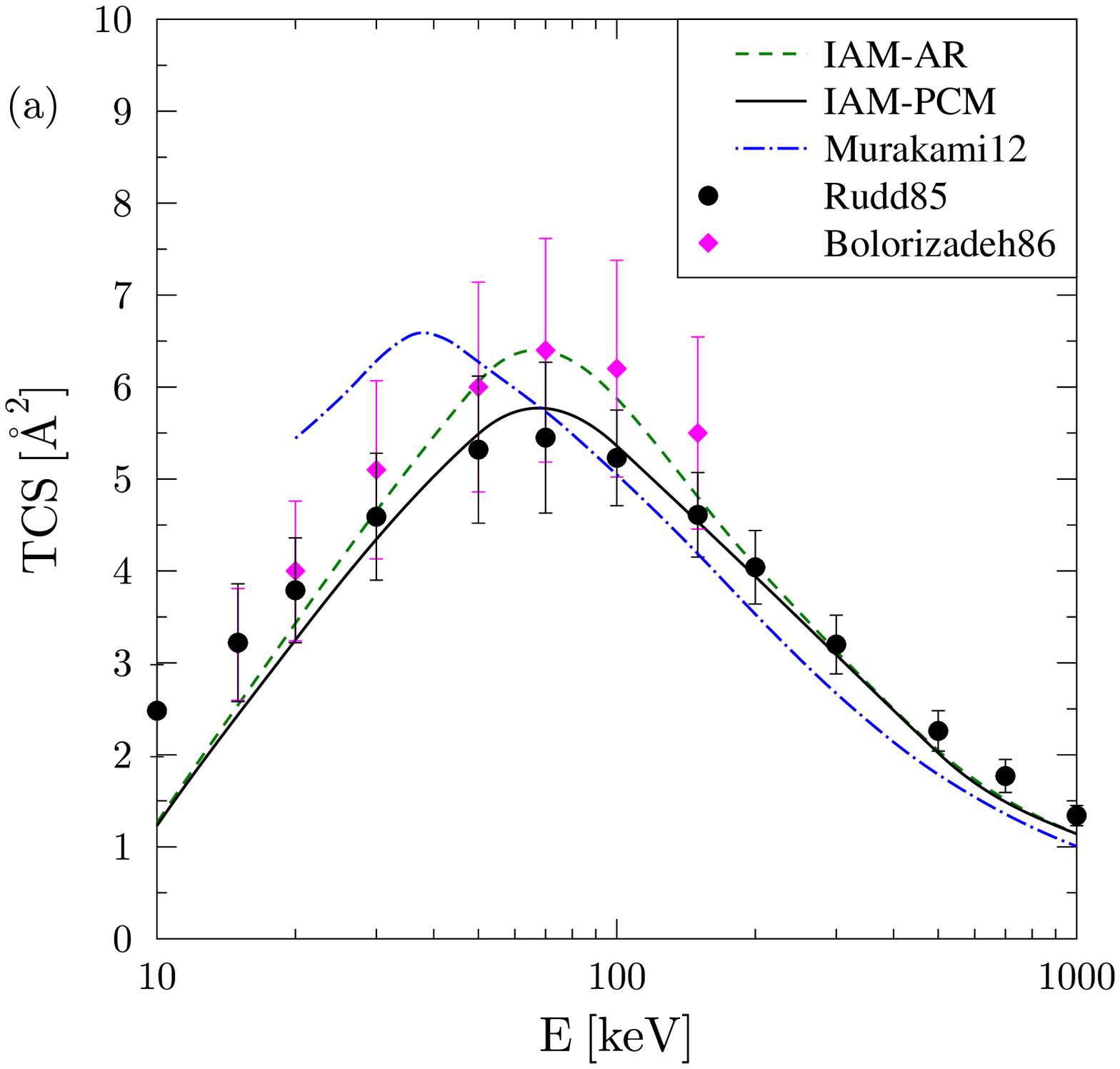}
}
\resizebox{0.49\textwidth}{!}{%
\includegraphics{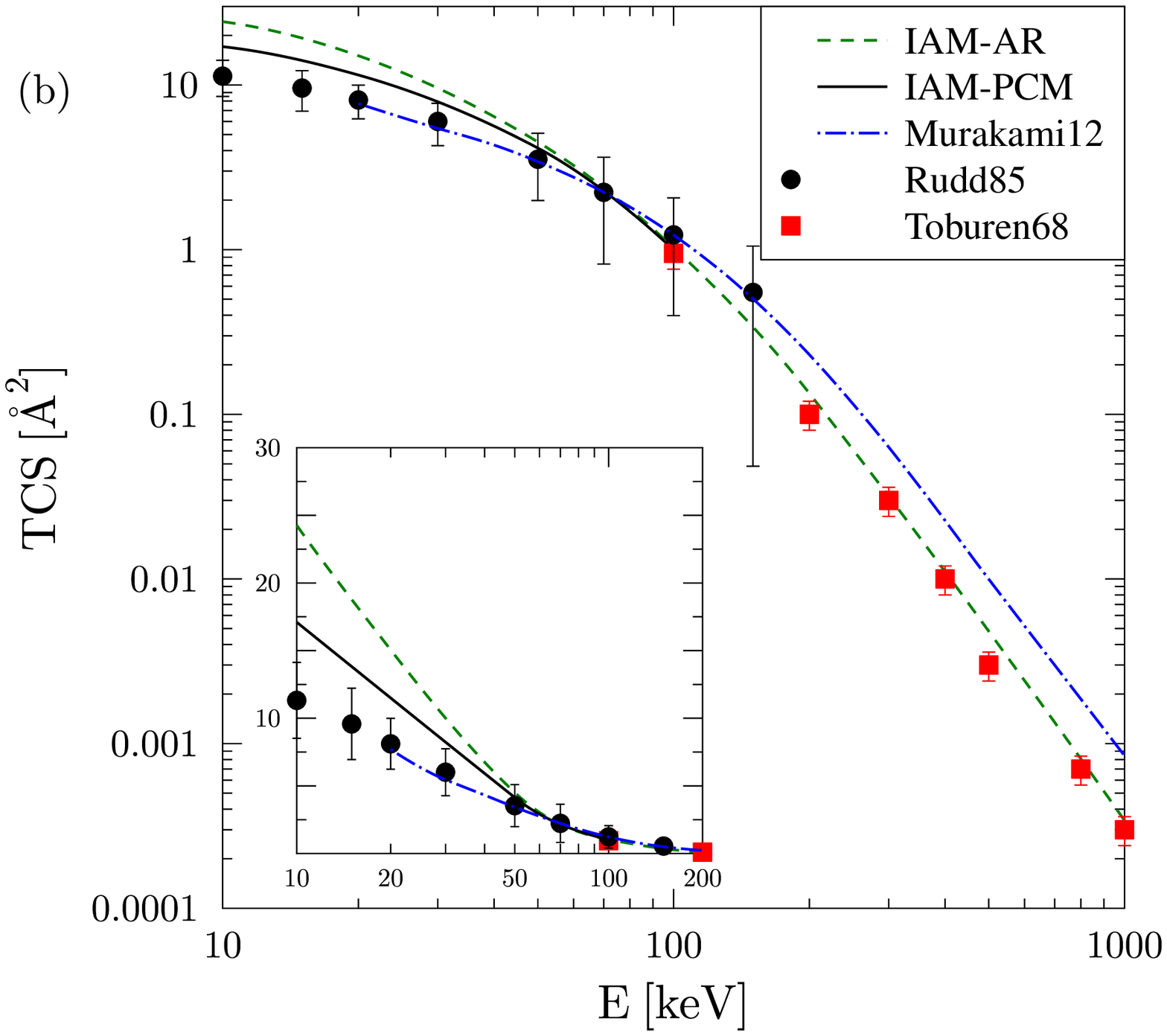}
}
\vspace{-2.5\baselineskip}
\caption{Total cross sections for (a) net ionization and (b) net capture
in p-H$_2$O collisions as functions of impact energy.
Murakami12 refers to the molecular TC-BGM calculation of reference \cite{mitsuko12}.
Experiments: Rudd85~\cite{Rudd85b}, Bolorizadeh86~\cite{Bolorizadeh86},
Toburen68~\cite{Toburen68}.}
\label{fig:h2o}
\end{center}
\end{figure*}

Figure~\ref{fig:100kev} shows net capture at the higher energy $E=100$ keV. At this
energy, the atomic net capture cross sections are small and the spheres do not overlap. The projection on the
$x$-$y$ plane is simply the sum of the atomic cross sections, i.e., the weight factors are equal to one 
and the IAM-AR result is recovered. 

In practice, we calculate the cross-sectional area of overlapping circular disks in the following way.
The $x$-$y$ plane is represented by a (pixel) matrix of dimension 
$1000 \times 1000$ with square elements
(pixels) whose size is determined by choosing a resolution 
(we typically use $0.01 \times 0.01$ {\AA}$^2$ pixels). 
The circular atomic cross section disks are 'colored'
according to their atomic identifier $j$ and the
pixel matrix is filled with the 
identifiers corresponding to the atomic cross sections from background to
foreground as seen by the impinging projectile. 
For each $j$ the area that is exposed to the projectile is 
determined by counting the visible pixels of that color
and the screening coefficients in equation (\ref{eq:scar-pcm}) 
are obtained by normalizing the
area to the total (unscreened) atomic cross section
\begin{equation}
  s_j^x(E,\alpha,\beta,\gamma) = \frac{\sigma_j^{{\rm vis} \, x} (E,\alpha,\beta,\gamma)}
                   { \sigma_j^{{\rm net} \, x} (E)} .
\label{eq:scar-coeff}
\end{equation}

It was noted in reference \cite{hjl16}
that the procedure can be criticized for overemphasizing the contribution of an atom
located at the front, while possibly completely neglecting the contribution of an atom
at the back of the molecule (cf. Figure~\ref{fig:10kev}b). 
However, as long as one is interested in net cross
sections only, this is a minor concern, since there is no need to attach physical
significance to the individual screening coefficients and
partial cross section areas. One may view them as purely auxiliary quantities  
to calculate the total projected area according to equation (\ref{eq:scar-pcm}). 
Obviously, the area can be  
decomposed in different, but equivalent ways.

To make contact with experimental data for randomly oriented molecules, IAM-PCM calculations are
carried out for a number of orientations and are averaged over the Euler angles.
For all results shown in this work
we exploit the fact that a rotation about the $z$-axis does 
not change the size of the visible area and vary only two out of three Euler angles 
on fine grids for a total of 40 $\times $ 40 orientations. 

As an illustration, we consider
the p-H$_2$O system in Figure~\ref{fig:h2o}.
We compare
IAM-PCM net capture and ionization cross sections with
experimental data and with previous TC-BGM calculations obtained in the molecular framework
mentioned in the Introduction,
in which simple self-consistent field wave functions were projected onto atomic
orbitals calculated in DFT \cite{mitsuko12}.

For net ionization (Figure~\ref{fig:h2o}a) the IAM-PCM outperforms the
molecular TC-BGM: The cross section
maximum appears at the correct position and the agreement with the 
measurements 
of Rudd and coworkers \cite{Rudd85b} is very good, except at energies below 20 keV where
these data are underestimated. 
By contrast, the molecular TC-BGM cross section curve peaks
at too low an energy and underestimates the experimental data above 100 keV.
The IAM-AR results show the same overall behaviour as IAM-PCM, 
except that the cross section values
are somewhat larger around the maximum, in seemingly excellent agreement with the
measurements of Bolorizadeh and Rudd \cite{Bolorizadeh86}. However, these cross
section data have relatively large error bars.
They were obtained from integrating absolute differential measurements 
and are deemed less accurate than those of reference \cite{Rudd85b},
which were obtained from a more direct parallel-plate-capacitor method. 
Overall, the comparison indicates that the inclusion of geometric screening
corrections via the IAM-PCM represents an improvement.

This becomes more obvious in the case of net capture. 
The linear plot in the inset of Figure~\ref{fig:h2o}b shows that the simple IAM-AR results in
a strong overestimation towards low energies where the atomic capture cross sections
are large (cf. Figure~\ref{fig:net-ao}). The overlap effect is significant (cf. Figure~\ref{fig:10kev})
and leads to a substantial reduction of the molecular cross section.
Still, the IAM-PCM results overestimate the experimental data at energies below 30 keV. 
It was argued in reference \cite{hjl16} that this overestimation is a consequence of the strong (resonant)
p-H contributions in the IAM, which are unphysical given that there is no resonant capture
channel in the p-H$_2$O collision system. The comparison with the molecular TC-BGM calculations
confirms this. Down to the lowest energy of 20 keV for which these calculations were carried out
they are in excellent agreement with the experimental data.

The situation is different at energies above 100 keV where the overlap effect in the IAM is
negligible (cf. Figure~\ref{fig:100kev}). IAM-PCM and IAM-AR results coincide and are in excellent
agreement with the measurements of Toburen {\it et al.} \cite{Toburen68}. The molecular TC-BGM
cross section is higher by about a factor of two in this region. 
No explanation
for this discrepancy has been found yet~\cite{kirchner13}.

\section{Results for proton-cluster collisions}
\label{sec:results}

Motivated by the goal to aid the microscopic understanding of proton cancer therapy
a recent theoretical work looked into proton collisions from water clusters at
$E=100$ keV~\cite{Privett17}. This is the region of the so-called Bragg peak,
which marks the point of maximum energy deposition near the end 
of the path of an ion traveling through matter~\cite{bichsel13}.

The calculation of reference \cite{Privett17} was based on the sim\-plest-level electron nuclear dynamics (SLEND) method
(see also reference~\cite{Privett14}),
in which classically moving nuclei are nonadiabatically coupled to
electrons represented in terms of an unrestricted Hartree-Fock (UHF) determinantal wave function.
Based on calculations for (H$_2$O)$_n$ with $n=1,\ldots , 6$ it was found that the total (one-electron) capture
cross section $\sigma(n)$ scaled as $n^{2/3}$. This was rationalized by associating each
cluster with a sphere of volume $V(n)$, assuming $V(n) \propto n$ and arguing that the 
capture cross section should be proportional to the effective area of the sphere 
exposed to the incident ion. Ionization was not considered in reference~\cite{Privett17}, 
since the basis sets used
did not allow for a representation of the continuum part of the spectrum. In
addition, ETFs (cf. equation~(\ref{eq:genbgm})) were neglected.

\begin{figure*}
\begin{center}
\resizebox{0.95\textwidth}{!}{%
\includegraphics{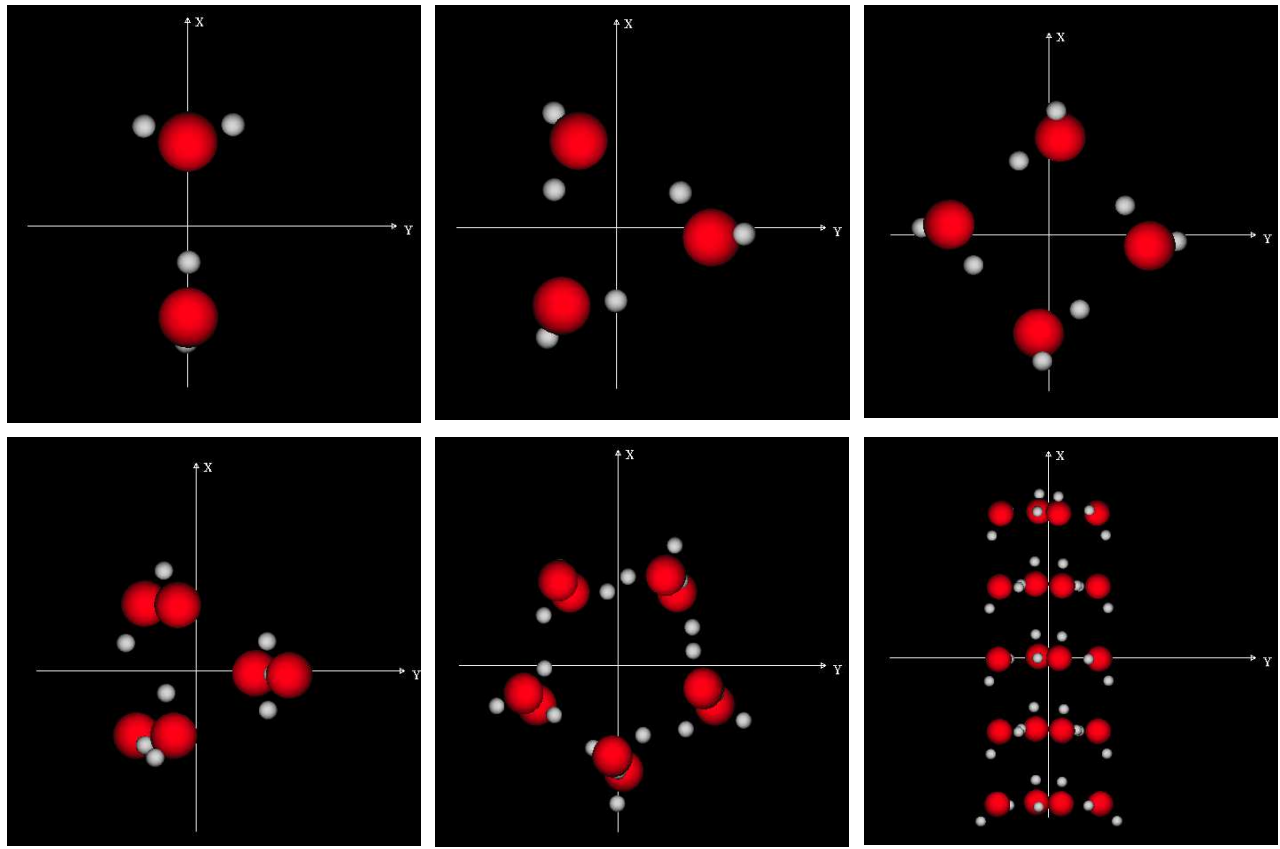}
}
\caption{Net capture cross sections
in p-(H$_2$O)$_n$ collisions at $E=100$ keV for $n=2,3,4,6,10,20$ from top left to
bottom right. The representation is analogous
to those of Figures~\ref{fig:10kev}b and~\ref{fig:100kev}b with radii   
determined acoording to equation~(\ref{eq:radius}). The corresponding plots
of net ionization are similar, but show larger disks and more significant overlap, since
the atomic ionization cross sections are larger.}
\label{fig:h2o-clusters}
\end{center}
\end{figure*}

\begin{figure}
\begin{center}
\resizebox{0.49\textwidth}{!}{%
\includegraphics{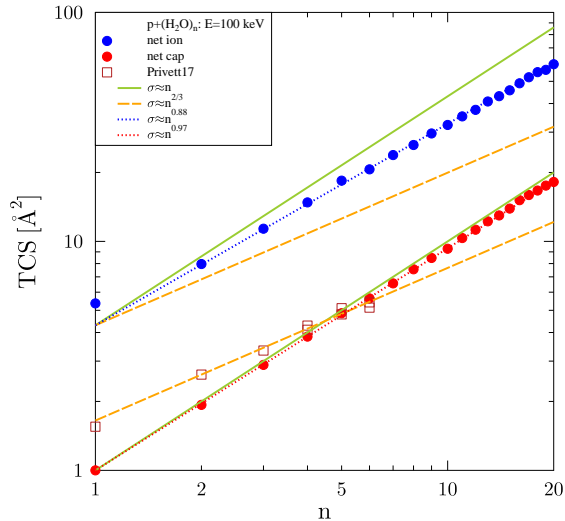}
}
\vspace{-3.5\baselineskip}
\caption{Total cross sections for net ionization and net capture
in p-(H$_2$O)$_n$ collisions at $E=100$ keV as functions of cluster size $n$. The straight lines
are obtained from equation~(\ref{eq:para1}) for different parameter choices and
are included to guide the eye. Privett17: SLEND calculation for one-electron capture
from reference~\cite{Privett17}.}
\label{fig:h2o-scale-100}
\end{center}
\end{figure}

We have applied the IAM-PCM to test the prediction of reference \cite{Privett17} and to explore the
scaling of both net capture and net ionization cross sections in p-(H$_2$O)$_n$ collisions
in the impact energy range from 10 to 1000 keV and for cluster sizes up to $n=20$. Specifically, we used the
set of isomers included in the Cambridge Cluster 
Database~\cite{CCD}\footnote{For (H$_2$O)$_6$ we chose the prism structure
and omitted the cage structure.}, 
whose structures were calculated at the restricted
Hartree-Fock/6-31G$(d,p)$ level \cite{Maheshwary01}.
Figure~\ref{fig:h2o-clusters} shows the IAM-PCM
net capture cross sections at $E=100$ keV for a subset of these clusters and arbitrary
geometries in a similar representation as used in Figures~\ref{fig:10kev}b and \ref{fig:100kev}b.
As a consequence of the relatively large distances between the monomers in the
clusters and the relative weakness of electron capture at 100 keV (cf. Figure~\ref{fig:h2o}b)
the overlap effect is small. This suggests the cross section scaling 
$\sigma^{{\rm net \, cap} }(n) \propto n^\alpha$ with a value of $\alpha$ close to one. 
Indeed, as Figure~\ref{fig:h2o-scale-100} shows, the IAM-PCM capture results for $n=1,\ldots ,20$
are almost perfectly fitted by
\begin{equation}
   \sigma^{{\rm net \,}x }(n) = a n^{\alpha}
\label{eq:para1}
\end{equation}
with $a=1.0$ {\AA}$^2$ and $\alpha=0.97$. Here $a$ represents an effective capture (or ionization)
cross section (in {\AA}$^2$) for the case $n=1$, but is treated as a fit parameter in order not to give
too much weight to the monomer. 

Ionization is stronger than capture at $E=100$ keV
(cf. Figure~\ref{fig:h2o}) and, accordingly, the overlap effect is larger. This translates into
the optimal fit parameter  $\alpha = 0.88$, which is still substantially larger than the value
$\alpha = 0.67$ found by Privett {\it et al}. \cite{Privett17}.
The different scaling behaviour between our calculations (which treat ionization properly)
and those of reference~\cite{Privett17} may have various reasons. Our calculations are based
on a model, whereas Privett {\it et al.} considered the molecular structure of the water clusters
in the UHF framework. As mentioned above, ETFs and ionization channels were neglected in their calculations.
Also, they did not consider net capture, but one-electron capture. The latter
is probably a minor concern given that both quantities should be similar in a calculation in which the
only other contribution to net capture is two-electron capture. 

\begin{figure}
\begin{center}
\resizebox{0.49\textwidth}{!}{%
\includegraphics{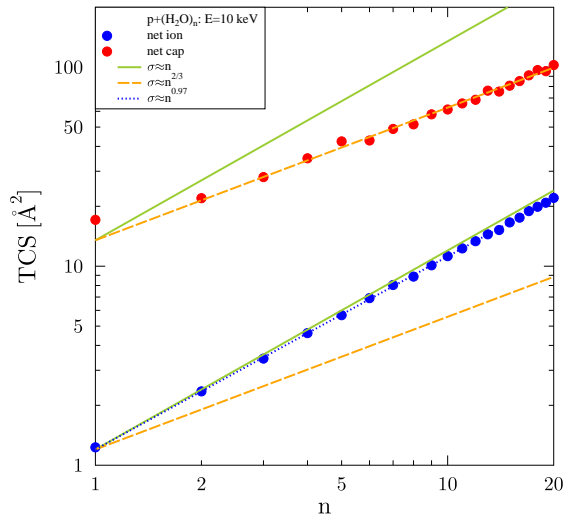}
}
\vspace{-3.5\baselineskip}
\caption{Total cross sections for net ionization and net capture
in p-(H$_2$O)$_n$ collisions at $E=10$ keV as functions of cluster size $n$. The straight lines
are obtained from equation~(\ref{eq:para1}) for different parameter choices and
are included to guide the eye.}
\label{fig:h2o-scale-10}
\end{center}
\end{figure}

Figure~\ref{fig:h2o-scale-10} shows IAM-PCM results for net capture and net ionization at $E=10$ keV.
For capture the overlap effect is large at low impact energy and the best fit of the calculations is obtained with $\alpha = 0.67$.
By contrast, ionization is weak and $\sigma^{{\rm net \, ion} }(n)$ scales almost linearly
with $n$. Linear scaling is also obtained at high energies where the IAM-PCM cross sections for 
net ionization and net capture approach the IAM-AR predictions. 

\begin{figure}
\begin{center}
\resizebox{0.49\textwidth}{!}{%
\includegraphics{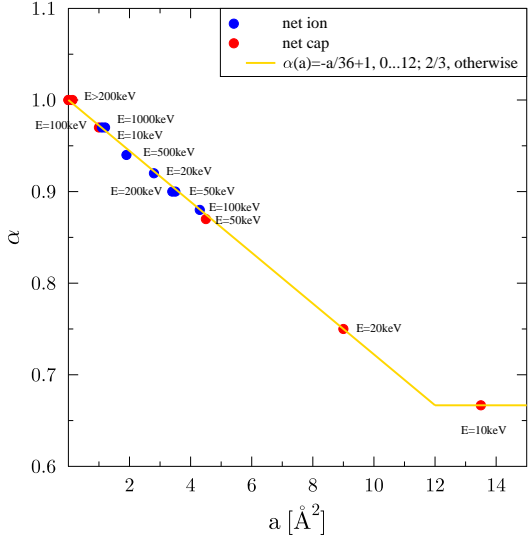}
}
\vspace{-3.5\baselineskip}
\caption{The exponent $\alpha$ in equation~(\ref{eq:para1}) for net ionization and net capture
in p-(H$_2$O)$_n$ collisions plotted versus the parameter $a$. Each data point corresponds to the best fit of the
IAM-PCM results for $\sigma^{{\rm net \,} x }(n)$ by 
equation~(\ref{eq:para1}) at the indicated impact energy. 
The full line corresponds to the parametrization~(\ref{eq:para2}).}
\label{fig:water-para}
\end{center}
\end{figure}

We tabulated the optimal parameters $\alpha$ and $a$ for both net ionization and net capture
at all impact energy values in the $10 \le E \le 1000$ keV
range for which we carried out calculations and found 
that the IAM-PCM cross section results
can be parametrized by using equation~(\ref{eq:para1})
and assuming
\begin{eqnarray}
\alpha(a)  &=&  
\left\lbrace
\begin{array}{ll}
-a/36.0 + 1.0  &  \; \mbox{if}\;\; a \le 12.0 \, {\rm \AA}^2 \\
  2/3 &  \; \mbox{otherwise} 
\end{array}
\right.
\label{eq:para2}
\end{eqnarray}
for the exponent.
This is demonstrated in Figure~\ref{fig:water-para}. Each point on the graph corresponds to
the best fit of the IAM-PCM $\sigma^{{\rm net} \, x }(n)$ results  
for a given $E$ to equation~(\ref{eq:para1}), i.e.,
to slope and intercept
of that straight line that fits the cross section results for capture or ionization
on a double-logarithmic
plot as used in Figures~\ref{fig:h2o-scale-100} and~\ref{fig:h2o-scale-10} for $E=100$ and $E=10$ keV, respectively.

The only deviation from the almost perfect linear dependence of
$\alpha$ on $a$ is observed for net capture at the lowest energy $E=10$ keV 
(i.e., the point at $a=13.5$ \AA$^2$), suggesting that $\alpha $ cannot
fall below 0.67. This lower limit is implemented explicitly in the parametrization 
by the piecewise definition of $\alpha(a)$
and seems plausible given the arguments provided by Privett {\it et al.} \cite{Privett17}
and the geometrical construction of the IAM-PCM cross section. In other words, the IAM-PCM
appears to be consistent with those arguments in the limit of strong overlap.
In the limit of weak overlap, the IAM-PCM approaches the IAM-AR prediction of a
linear cross section scaling with cluster size $n$. 
Given the energy dependence of the atomic cross section magnitudes and overlaps 
the $n$-dependence is not universally determined by the
geometry of the cluster as the arguments provided by Privett {\it et al.} might suggest.

\begin{figure}
\begin{center}
\resizebox{0.49\textwidth}{!}{%
\includegraphics{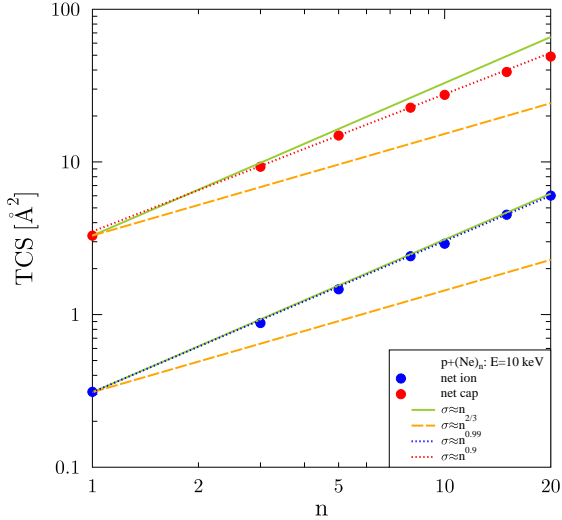}
}
\vspace{-3.5\baselineskip}
\caption{Total cross sections for net ionization and net capture
in p-Ne$_n$ collisions at $E=10$ keV as functions of cluster size $n$. The straight lines
are obtained from equation~(\ref{eq:para1}) for different parameter choices and
are included to guide the eye.}
\label{fig:ne-scale}
\end{center}
\end{figure}

To further test 
these observations we carried out IAM-PCM calculations for proton impact on 
neon clusters. The relevant structure information is also taken from the Cambridge Cluster
Database using $d=3.3$ a.u. as the internuclear distance of the dimer~\cite{CCD}. 
For p-Ne$_n$ collisions with $n=1, \ldots ,20$ we find $\alpha \ge 0.9$ for both net
capture and net ionization in the entire impact
energy range from 10 to 1000 keV.
Figure~\ref{fig:ne-scale} shows the IAM-PCM cross sections and the fits according to equation~(\ref{eq:para1}) 
at $E=10$ keV. 
The cross sections are smaller than for p-(H$_2$O)$_n$ collisions, since the Ne electrons are
more tightly bound. 
Given that the average distance between the monomers is similar in neon and water clusters, the atomic
cross section overlaps are smaller and $\alpha$ is larger for the former.

\begin{figure}
\begin{center}
\resizebox{0.49\textwidth}{!}{%
\includegraphics{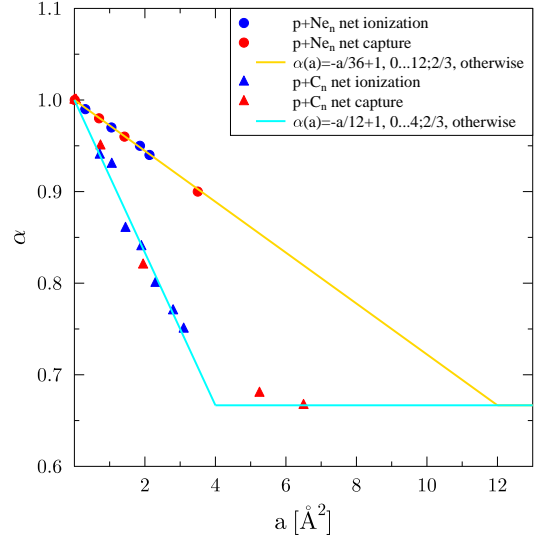}
}
\vspace{-3.5\baselineskip}
\caption{The exponent $\alpha$ in equation~(\ref{eq:para1}) for net ionization and net capture
in p-Ne$_n$ and p-C$_n$ collisions plotted versus the parameter $a$. Each data point corresponds to the best fit of the
IAM-PCM results for $\sigma^{{\rm net \,} x }(n)$ by 
equation~(\ref{eq:para1}) at a given impact energy. 
The yellow line corresponds to the parametrization~(\ref{eq:para2}) and
the light-blue line to (\ref{eq:para3}).}
\label{fig:ne-para}
\end{center}
\end{figure}

Remarkably, the p-Ne$_n$ results over the entire impact energy range
can also be parametrized by equation~(\ref{eq:para2}). This is shown in Figure~\ref{fig:ne-para}, which is analogous to Figure~\ref{fig:water-para} for
p-(H$_2$O)$_n$ collisions. The range of $\alpha(a)$ points for neon clusters 
is compressed compared to Figure~\ref{fig:water-para}
reflecting the smaller atomic cross sections and overlaps. 

\begin{figure}
\begin{center}
\resizebox{0.49\textwidth}{!}{%
\includegraphics{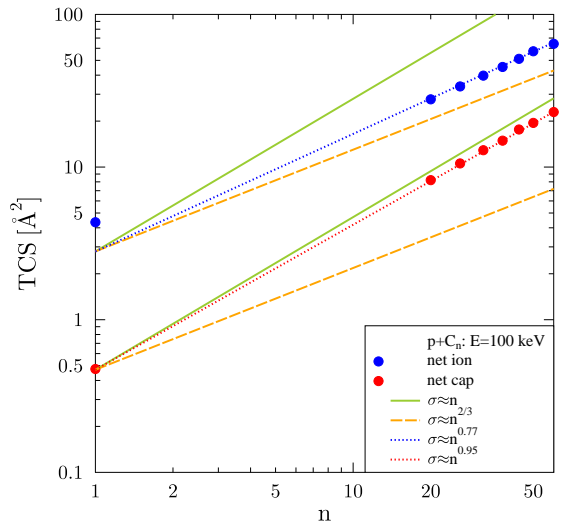}
}
\vspace{-3.5\baselineskip}
\caption{Total cross sections for net ionization and net capture
in p-C$_n$ collisions at $E=100$ keV as functions of cluster size $n$. The straight lines
are obtained from  equation~(\ref{eq:para1}) for different parameter choices and
are included to guide the eye.}
\label{fig:carbon-scale}
\end{center}
\end{figure}

Finally, we consider proton collisions from a
selection of carbon clusters C$_n$ with $20 \le n \le 60$. 
The relevant
structure information is taken from reference~\cite{carbonstructure}. 
Similarly to the (H$_2$O)$_n$ case
we find that net capture
scales as $n^{2/3}$ at low energy, while $\alpha$ approaches unity more slowly 
than for water clusters towards higher energies.
In the case of net ionization we also 
obtain somewhat smaller $\alpha$ values for C$_n$ than for (H$_2$O)$_n$ signaling larger overlaps.
Figure~\ref{fig:carbon-scale} illustrates these observations for $E=100$ keV. For capture the
optimal $\alpha$ value is 0.95, while for ionization $\alpha = 0.77$ provides the best fit of the
IAM-PCM calculations. This is to be contrasted with $\alpha = 0.97$ and $\alpha = 0.88$ for
p-(H$_2$O)$_n$ collisions, respectively (cf. Figure~\ref{fig:h2o-scale-100}).

The parametrization (\ref{eq:para2}) does not work for fullerenes, but we found that the 
ansatz (\ref{eq:para1}) together with
\begin{eqnarray}
\alpha(a)  &=&  
\left\lbrace
\begin{array}{ll}
-a/12.0 + 1.0  &  \; \mbox{if}\;\; a \le 4.0  \, {\rm \AA}^2 \\
  2/3 & \; \mbox{otherwise} 
\end{array}
\right.
\label{eq:para3}
\end{eqnarray}
provides a good fit of the results in the 10 to 1000 keV impact energy range.
These results are included in Figure~\ref{fig:ne-para}.
One may argue that the slope of $\alpha (a)$ for a given cluster species
is reflective of the
average distance between monomers in the clusters. Additional calculations for p-Ar$_n$
support this and all data taken together suggest that the slope is approximately inversely proportional
to that distance. 
Systematic measurements for a set of clusters over a range of impact energies 
would be highly desirable to test these predictions.

\begin{figure}
\begin{center}
\resizebox{0.49\textwidth}{!}{%
\includegraphics{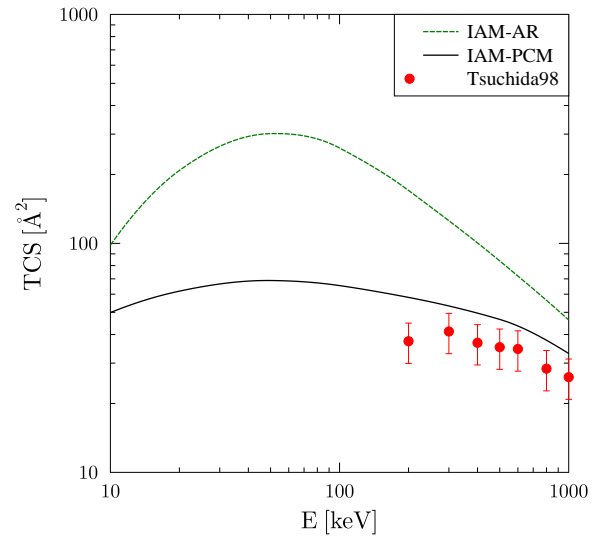}
}
\vspace{-3.5\baselineskip}
\caption{Total cross section for net ionization 
in p-C$_{60}$ collisions as function of impact energy. Experimental data: Tsuchida98 \cite{Tsuchida98}.}
\label{fig:c60}
\end{center}
\end{figure}

Experimental data
are available for net ionization of C$_{60}$ at high impact energies \cite{Tsuchida98}. In Figure~\ref{fig:c60}
we compare these measurements with IAM-PCM and IAM-AR calculations. The overlap effect is significant,
reducing the net ionization cross section by more than a factor of two for most of the impact
energy interval shown. The experimental data are even lower than the IAM-PCM results
with the latter just lying outside of the error bars. 
One can regard the agreement as fair. Clearly, data at lower energies (and for net capture as well)
would be needed for a better assessment
of the quality of the IAM-PCM results.

\section{Concluding remarks}
\label{sec:conclusions}
34 years after the publication of the Runge-Gross theorem full-fledged TDDFT
calculations for ion-impact collisions
have remained a rarity compared to the widespread application of TDDFT
to laser-matter interaction problems. 
As yet, simplified
approaches and models are indispensable for a semi-quantitative understanding
of electron removal processes in collisions involving complex multicenter Coulomb systems. 
The IAM-PCM is one such model. It is based on a geometrical interpretation of the
cross section as an effective area composed of overlapping circular disks whose areas represent
the atomic cross sections that contribute to net capture or net ionization in the system
of interest.  
The atomic cross sections are calculated based on a
TDDFT-inspired single-particle description using atomic ground-state DFT
potentials and the two-center basis generator method for orbital propagation. The
effective area calculation is carried out using a pixel counting method.

The IAM-PCM is flexible and efficient. Once the atomic cross sections have been calculated
and the required information on the geometric structure of the
target, i.e., the equilibrium positions of the nuclei, is available it takes about three
minutes on a single-core desktop or laptop
computer to calculate the net ionization or net capture cross section at a given
impact energy for a system as complex as C$_{60}$. 

To date, we have applied the IAM-PCM to proton collisions from 
a variety of targets: covalently bound molecules in reference~\cite{hjl96} and, in this work,
clusters with hydrogen bonds ((H$2$O)$_n$), van der Waals clusters (Ne$_n$), and
covalently-bound fullerenes (C$_n$). One major objective of this work has been
to test and generalize a scaling law found by Privett {\it et al.} \cite{Privett17}
in capture from water clusters at $E=100$ keV to capture and ionization over a wide
range of energies.  

Our results can be summarized as follows: Both net capture and net ionization cross
sections at a given impact energy scale as $n^\alpha$, but $\alpha$ varies as a 
function of $E$ and reaches the value of 2/3 found by Privett {\it et al.} for one-electron
capture only in
situations in which the atomic cross section overlaps are large. 
This is the case for capture at low impact energy ($E=10$ keV) in p-(H$2$O)$_n$
and p-C$_n$ collisions, but not in p-Ne$_n$ collisions and never for ionization.
For capture from water clusters at $E=100$ keV we find $\alpha = 0.97$ in stark
contrast to the result of Privett {\it et al.}

Furthermore, we showed that the variations of $\alpha$ can be modeled 
by the ansatz $\sigma^{{\rm net \,}x } = a n^{\alpha(a)}$ and a linear
functional dependence of $\alpha$ on $a$. Our results suggest that the slope
of this linear function is largely determined by the average distance between the monomers
in a given cluster. It will be interesting to see how general this result is and
where its limitations are. Further calculations for other systems
and, more importantly, systematic experimental measurements will be required to answer this question.

%\section*{Authors contributions}
%All the authors were involved in the preparation of the manuscript.
%All the authors have read and approved the final manuscript.

%\begin{acknowledgment}
\vspace{1\baselineskip}
\noindent
This work was supported by the Natural Sciences and Engineering Research Council of Canada (NSERC).
One of us (H. J. L.) would like to thank the Center for Scientific Computing, University of Frankfurt 
for making their
High Performance Computing facilities available.
%\end{acknowledgment}

%
% BibTeX users please use
%\bibliographystyle{unsrt}
\bibliography{pcm-epjb}
%\begin{thebibliography}{}
%Author, \textit{Book title} (Publisher, place year) page numbers
% etc
%\end{thebibliography}

\end{document}